\documentclass[aps,prb,preprint,superscriptaddress,longbibliography]{revtex4-2}
\usepackage[dvipdfmx]{graphicx}
\usepackage{mathrsfs}
\usepackage{url}
\newcommand{\tc}{$T_{\rm c}$}
\newcommand{\tl}{$1/T_1$}

\newcommand{\tlt}{$1/T_1T$}
\newcommand{\nbr}{NbRh$_2$B$_2$}
\newcommand{\tar}{TaRh$_2$B$_2$}
\newcommand{\xr}{NbRh$_2$B$_2$ and TaRh$_2$B$_2$}

\begin{document}
\title{Antiferromagnetic spin fluctuations and superconductivity in NbRh$_2$B$_2$ and TaRh$_2$B$_2$ with a chiral crystal structure}

\author{Kazuaki Matano}
\affiliation{Department of Physics, Okayama University, Okayama 700-8530, Japan}

\author{Ryo Ogura}
\affiliation{Department of Physics, Okayama University, Okayama 700-8530, Japan}

\author{Mateo Fountaine\footnote{Present address: \'{E}cole normale sup\'{e}rieure Paris-Saclay, Cachan Cedex 94235, France}}
\affiliation{Department of Physics, Okayama University, Okayama 700-8530, Japan}

\author{Harald O. Jeschke}
\affiliation{Research Institute for Interdisciplinary Science, Okayama University, Okayama 700-8530, Japan}

\author{Shinji Kawasaki}
\affiliation{Department of Physics, Okayama University, Okayama 700-8530, Japan}

\author{Guo-qing Zheng}
\affiliation{Department of Physics, Okayama University, Okayama 700-8530, Japan}

\date{\today}

\begin{abstract}
We report the $^{11}$B nuclear magnetic resonance (NMR) measurements on non-centrosymmetric superconductors NbRh$_2$B$_2$ (superconducting transition temperature $T_c$ = 7.8 K) and TaRh$_2$B$_2$ ($T_c$ = 5.9 K) with a chiral crystal structure. 
The nuclear spin-lattice relaxation rate $1/T_1$ shows no coherence peak below $T_{\rm c}$,
which suggests unconventional nature of the superconductivity.
In the normal state, $1/T_1T$ increases with decreasing temperature $T$ at low temperatures below $T$ = 200 K for TaRh$_2$B$_2$ and $T$ = 15 K for NbRh$_2$B$_2$, while the Knight shift remains constant. These results suggest the presence of antiferromagnetic spin fluctuations in both compounds. The stronger spin fluctuations in TaRh$_2$B$_2$ compared to NbRh$_2$B$_2$ is discussed in the context of spin-orbit coupling.
\end{abstract}

\date{\today}

\maketitle

\section{introduction}
For a long time, the relationship between strong electronic correlations and unconventional superconductivity has been a major theme in condensed matter physics\cite{RevModPhys.78.17,Moriya_review_cuprate_doi:10.1080/000187300412248,cuprate_Monthoux_PhysRevB.49.4261}. Antiferromagnetic spin fluctuations due to $3d$ electrons are essential for the high-temperature superconductivity in copper oxides\cite{cuprate_Monthoux_PhysRevB.49.4261,cuprate_NMR_Ohsugi_doi:10.1143/JPSJ.63.700}, and the same is true in iron pnictides\cite{Mazin_FeAs_theory_PhysRevLett.101.057003,PhysRevLett.101.087004,Matano_BaFeAs,PhysRevLett.108.047001} and cobalt-based superconductors\cite{Matano_NaxCoO2_EPL}. In contrast, 4$d$ and 5$d$ electrons are usually not as strongly correlated as 3$d$ electrons because of the considerably larger spatial extension of the wavefunction, while spin-orbit coupling (SOC) is stronger.

Superconductivity in compounds containing 4$d$ or 5$d$ elements has also attracted attention in recent years, as spin-triplet and spin-singlet mixed superconductivity has been discovered in some non-centrosymmetric superconductors (NCSs) containing 5$d$ elements\cite{Nishiyama_PhysRevLett.98.047002,PhysRevLett.97.017006,Harada_PhysRevB.86.220502}. Such a mixed state is explained by the antisymmetric spin-orbit coupling (ASOC) interaction \cite{PhysRevLett.87.037004,PhysRevLett.92.097001,1367-2630-6-1-115}, and can show topological properties\cite{sato_PhysRevB.79.094504,tanaka_PhysRevLett.105.097002}. Broken inversion symmetry and strong SOC also lead to the realization of Weyl semimetals\cite{Weyl_PhysRevX.5.011029,YangLiuSunEtAl2015}, where a huge orbital diamagnetic response has been found \cite{TaAs_PhysRevB.101.241110} and an unconventional superconducting state  has been suggested \cite{WeylSC_PhysRevB.86.054504,WeylSC_PhysRevB.86.214514,MoTe2_QiNaumovAliEtAl2016}.

Among NCSs, only Li$_2$Pt$_3$B\cite{Nishiyama_PhysRevLett.98.047002} and CePt$_3$Si\cite{PhysRevLett.100.107003} show pronounced signatures of spin-triplet properties. Li$_2$Pt$_3$B is a weakly correlated metal, but CePt$_3$Si is a heavy fermion superconductor and electron correlation is strong. After Li$_2$Pt$_3$B was reported, a number of NCSs with weak electron correlations were found\cite{Mg10Ir19B16,doi:10.1143/JPSJ.72.1724,GOLL20081065,BiPd,PhysRevB.89.020505}, but no spin-triplet superconductivity was reported\cite{Tahara_PhysRevB.80.060503,PhysRevB.82.064511,matano_JPSJ.82.084711,SinghHillierMazidianEtAl2014,Matano_Re6Zr_PhysRevB.94.214513,Maeda_PbTaSe2_PhysRevB.97.184510}. The reason why spin-triplet superconductivity was not observed in those compounds was probably because the ASOC was not large enough\cite{Harada_PhysRevB.86.220502}. The question then, is how to enhance the ASOC. It has been found that the enhancement of the ASOC in Li$_2$(Pd$_{1-x}$Pt$_x$)$_3$B is caused by the decrease in the angle between B(Pd,Pt)$_6$ octahedra, which enhances the breaking of the spatial inversion symmetry\cite{Harada_PhysRevB.86.220502}. Inspection of the evolution of Li$_2$(Pd$_{1-x}$Pt$_x$)$_3$B with $x$ reveals that the local distortion in the crystal structure is another important factor in addition to the presence of the heavier element Pt\cite{Harada_PhysRevB.86.220502}. Note that Li$_2$(Pd,Pt)$_3$B has a chiral crystal structure, which is an advantage for achieving a large extent of symmetry breaking.

{\xr} are recently discovered superconductors ({\tc} = 7.8 K for {\nbr}, 5.8 K for {\tar}) with a chiral crystal structure\cite{tarh2b2_discover}. They have a structure with space group $P3_1$ and a large upper critical field ($H_{c2}$ = 18.0 T for {\nbr}, 11.7 T for {\tar}) exceeding the Pauli limit\cite{tarh2b2_discover,Mayoh_2019}.  It is interesting to note the similarity of the relationship between {\xr} to that between Li$_2$Pd$_3$B and Li$_2$Pt$_3$B. Like Pd and Pt, Nb and Ta are located in the same group in the periodic table. Therefore, they provide a good platform to study the interplay between ASOC, electron correlations and superconductivity.

Here we report $^{11}$B nuclear magnetic resonance (NMR) measurements of polycrystalline samples of {\xr}. We find an increase of {\tlt} with decreasing temperature at low temperatures for both compounds. The Knight shift is constant in the temperature range where {\tlt} is increased. These results suggest the existence of antiferromagnetic spin correlations. However, the magnitude of spin correlations is different, and it is more significant for {\tar} where the SOC is larger. In the superconducting state, the spin-lattice relaxation rate {\tl} dropped below {\tc} without a coherence peak witch suggests unconventional superconductivity.

\section{experimental and theoretical}
\subsection{Sample preparation and characterization}
The polycrystalline samples of {\xr} were synthesized by heating a mixture of Nb(Ta), Rh and B in a vacuum. Elemental Nb (99.99\%), Ta (99.9\%), Rh (99.9\%), and B (99\%) were used. Powders of the starting materials Ta/Nb, Rh, and B were weighed in a ratio of 1:1.9:2.1, crushed using a mortar and pestle, and pressed into a pellet. The pellets were wrapped in Ta foil and heated at 1200 \(^\circ\)C for 6 h while vacuuming in a quartz tube with pressure on the order of 10$^{-1}$ Pa. The pellets were crushed into powder for X-ray and NMR measurements. The $T_c$ was determined by measuring the ac susceptibility using the {\it in situ} NMR coil. Dc susceptibility measurements were performed using a superconducting quantum interference device (SQUID) with the vibrating sample magnetometer (VSM).

\subsection{NMR measurements}
A standard phase-coherent pulsed NMR spectrometer was used to collect data. The NMR measurements were performed at an applied magnetic field $H_0$ = 3.0378 T. The nuclear gyromagnetic ratio $\gamma$=13.66 MHz/T was used for calculation of the Knight shift. The nuclear spin-lattice relaxation rate {\tlt} was measured by using a single saturation pulse. The spin-lattice relaxation time $T_1$ was measured by using a single saturating pulse. The recovery curves of the nuclear magnetization in all temperature ranges were fitted by a single stretched exponential function
\begin{eqnarray}
\frac{M_0-M(t)}{M_0} = 0.9 e^{\left(-\frac{6t}{T_1}\right)^\beta}+ 0.1e^{\left({-\frac{t}{T_1}}\right)^\beta}
\end{eqnarray}
to extract {\tl}, where $M_0$ is the nuclear magnetization in the thermal equilibrium, $M(t)$ is the nuclear magnetization at a time $t$ after the saturating pulse, and $\beta < 1$. In this study, $\beta$ is within the range of 0.6 to 0.8.

\subsection{Band calculations}
We used density functional theory (DFT) calculations within the full potential local orbital (FPLO) basis~\cite{Koepernik1999}. We use the generalized gradient approximation (GGA) to the exchange correlation functional~\cite{Perdew1996}, and we perform fully relativistic calculations in order to include the effects of spin-orbit (SO) coupling. For {\tar}, we use the crystal structure given in Ref.~\onlinecite{tarh2b2_discover}. For {\nbr} which is isostructural to {\tar}, only lattice parameters are given in Ref.~\onlinecite{tarh2b2_discover}. We use GGA calculations to optimize the internal coordinates of {\nbr}. We are confident that the DFT structure prediction works very well in these materials because application of the same relaxation to {\tar} leads only to a 60\,meV lower energy per formula unit and nearly unchanged atomic positions. We use $24\times 24\times 24$ $k$ meshes to calculate the band structures and densities of states. 

\section{results and discussion}
\subsection{Spin correlations in the normal state}
\begin{figure}[htbp]
\includegraphics[clip,width=80mm]{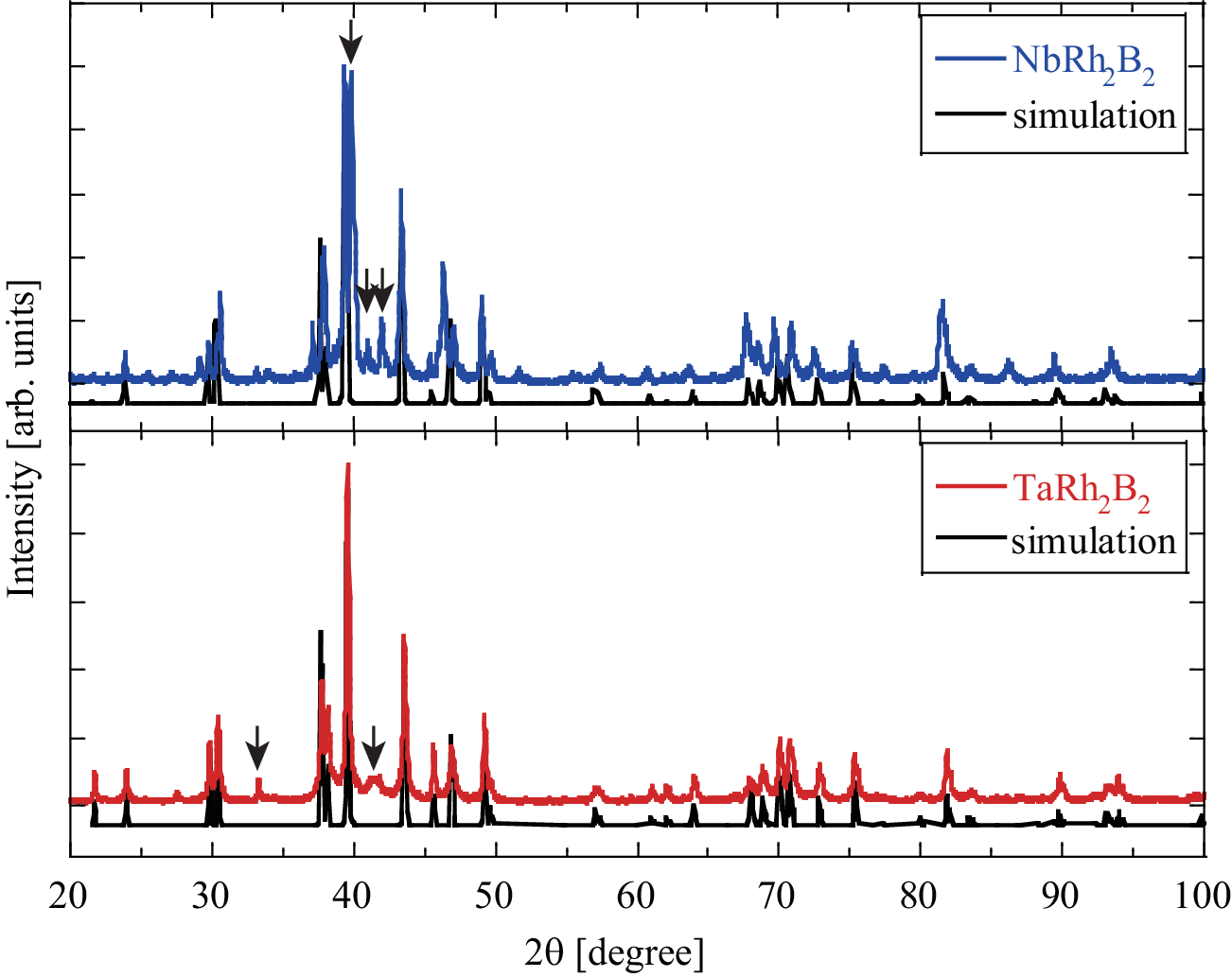}
\caption{\label{X-ray}(color online) XRD patterns for {\xr}. The theoretical curves(simulations) are obtained by the Rietveld method. Arrows indicate unidentified peaks.
}
\end{figure}
\begin{figure}[htbp]
\includegraphics[clip,width=80mm]{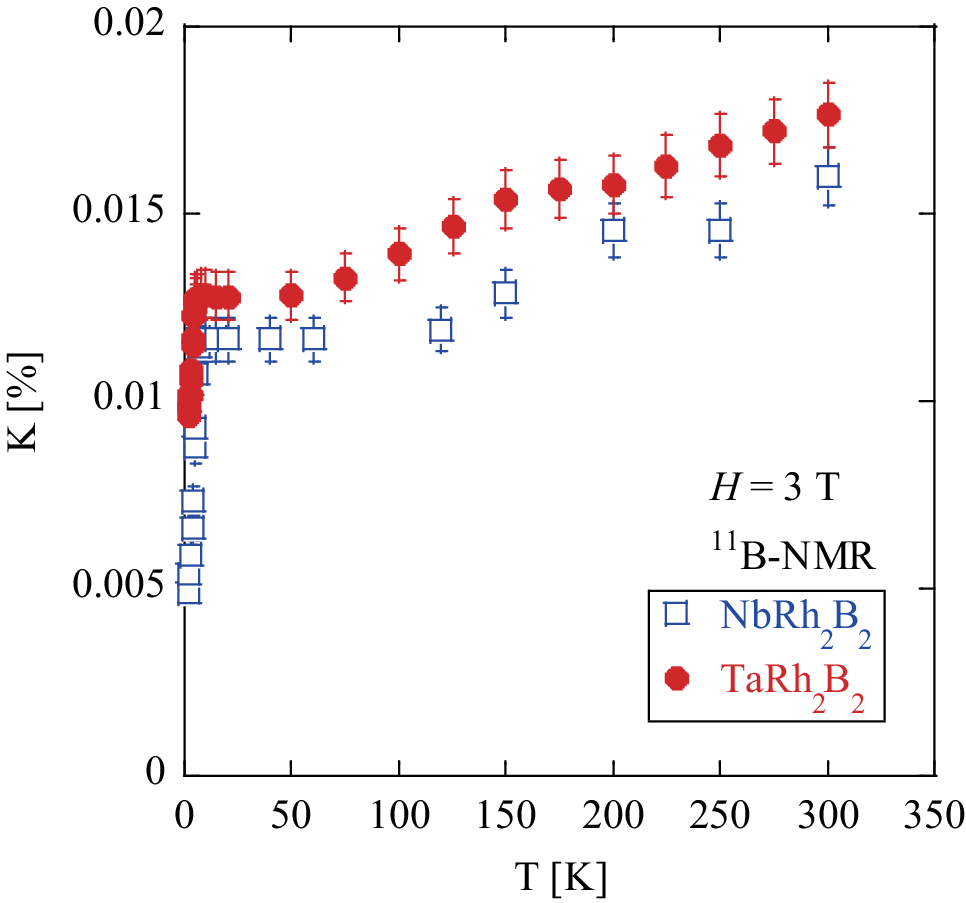}
\caption{\label{K}(color online) Temperature dependence of the Knight shift for {\xr}.}
\end{figure}

Figure \ref{X-ray} shows the powder X-ray diffraction (XRD) patterns for {\xr}. The results are generally in agreement with the simulations using the Rietveld method, but impurity peaks, which have also been observed in previous studies\cite{tarh2b2_discover}, were observed in both samples. 
\begin{figure}[htbp]
\includegraphics[clip,width=80mm]{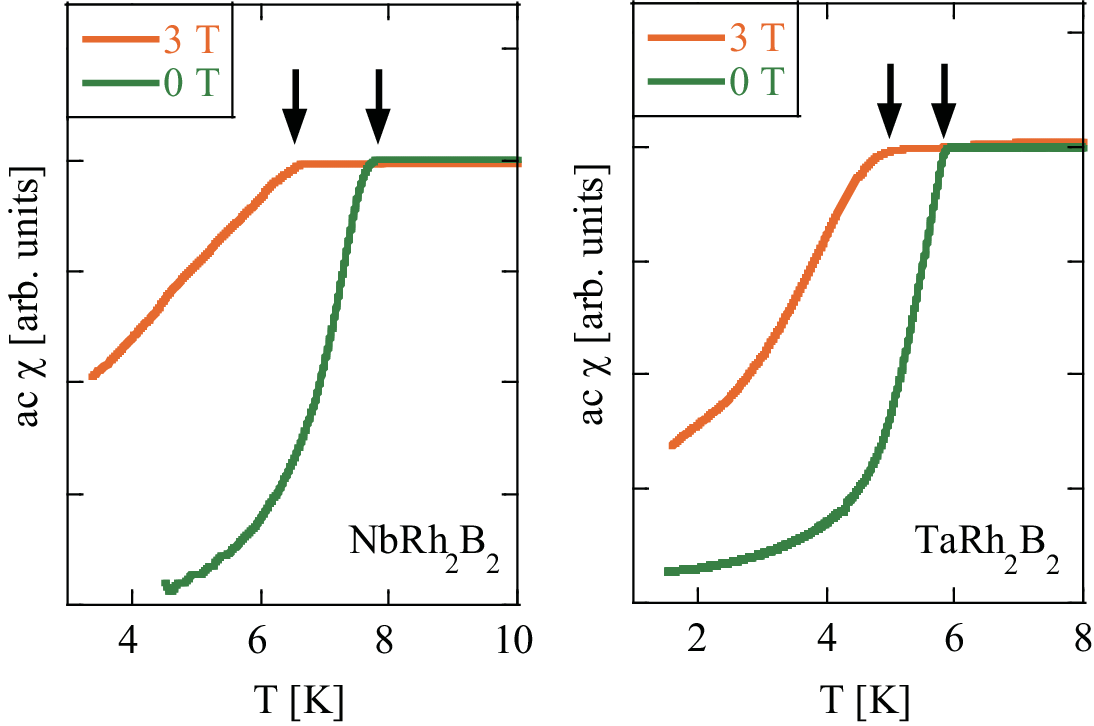}
\caption{\label{ac-chi}(color online) Ac susceptibility measured using the $in$ $situ$ NMR coil at zero and finite magnetic field (3 T). The arrows indicate {\tc} for each sample at different fields.}
\end{figure}
Figure \ref{ac-chi} shows the ac susceptibility measured using the {\it in situ} NMR coil. The {\tc} was determined as the onset temperature of the appearance of diamagnetism. The {\tc} for zero magnetic field is 7.6 K for {\nbr} and 5.8 K for {\tar}, respectively. When a magnetic field of 3 T was applied, {\tc} is reduced to 6.5 K for {\nbr} and 5.0 K for {\tar}, respectively. The obtained {\tc} is consistent with the previous report\cite{tarh2b2_discover}.
\begin{figure}[htbp]
\includegraphics[clip,width=80mm]{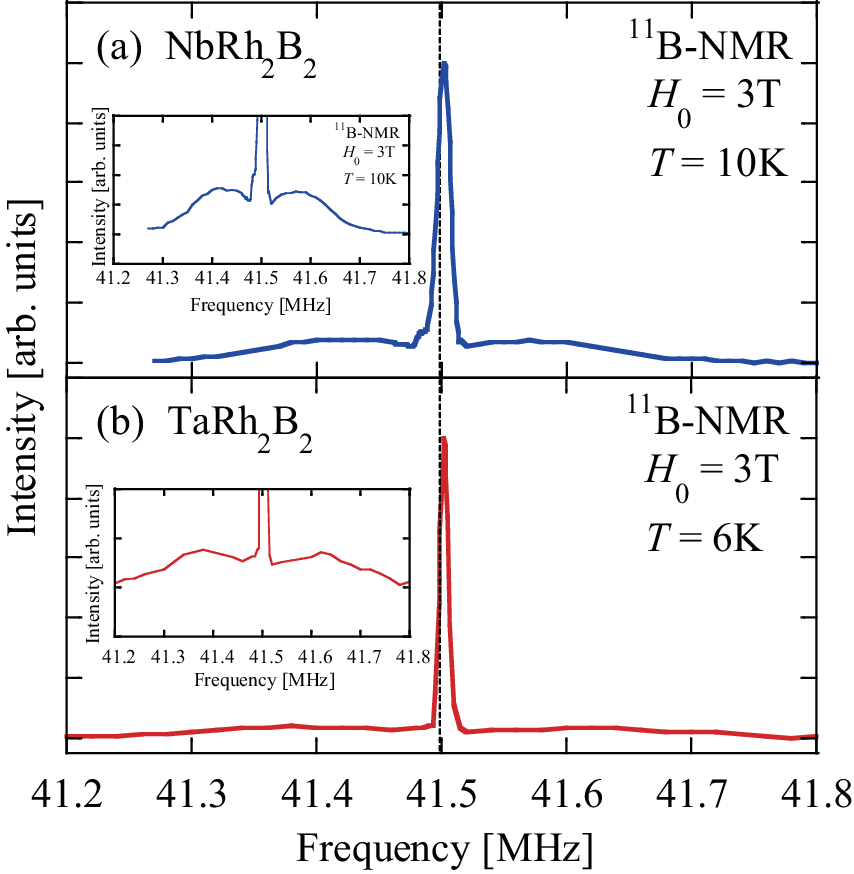}
\caption{\label{spec} (color online) 
(a) $^{11}$B-NMR spectra at 3.0 T and 6 K for {\nbr}.
(b) $^{11}$B-NMR spectra at 3.0 T and 10 K for {\tar}.
The dotted line shows the position of $K$=0.
}
\end{figure}
Figure \ref{spec} shows the $^{11}$B($I$ = 3/2)-NMR spectra for {\xr}. A typical powder pattern is observed for both samples. The full width at half maximum (FWHM) of the central peak is 10.2 kHz for {\nbr} and 7.4 kHz for {\tar}.
Figure \ref{K} shows the Knight shift, $K$, as a function of temperature. In both samples, $K$ decreases gradually toward lower temperatures and becomes constant at 60 K. Then, with the superconducting transition, $K$ decreased sharply with the onset of superconductivity. 
\begin{figure}[htbp]
\includegraphics[clip,width=80mm]{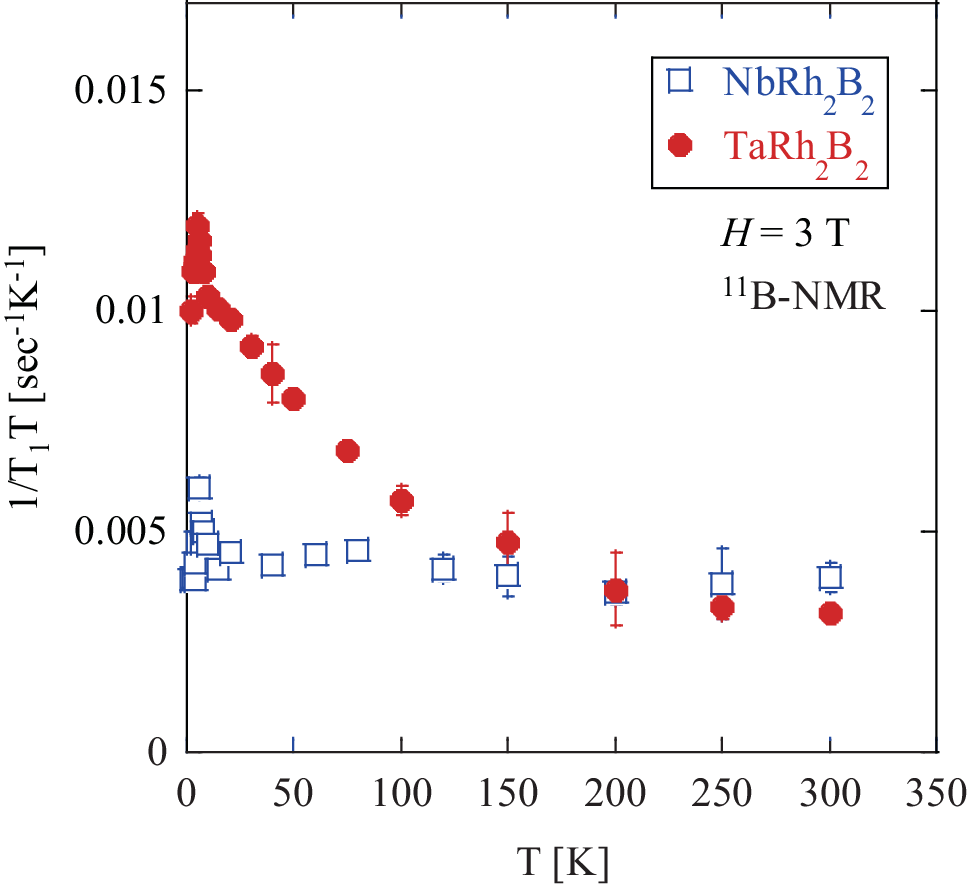}
\caption{\label{T1T}(color online) Temperature dependence of $1/T_1T$ measured at center peak for {\xr}.}
\end{figure}
Figure \ref{T1T} shows the temperature dependence of $1/T_1T$ for {\xr}. Above 200 K, $1/T_1T$ is constant for both compounds. However, at low temperatures, {\tlt} behaves differently. In {\tar}, {\tlt} is greatly enhanced below 200 K, while the enhancement is weak in {\nbr}.


In a general form, $1/T_1T$ is expressed as
\begin{eqnarray} 
\frac{1}{T_1T}= \frac{\pi k_{\rm B} \gamma^2_n }{(\gamma_e \hbar )^2} \sum_q A_{hf}^2 \frac{\chi ''_{\perp}(q,\omega)}{\omega}\,,
\end{eqnarray}
where $\chi ''_{\perp}(q,\omega)$ is the imaginary part of the dynamical susceptibility perpendicular to the applied field, and $\omega$ is the NMR frequency.
If one assumes that there is a peak around a finite wave vector $Q$ (due to spin fluctuation), then one may have the following approximation,
\begin{eqnarray} \nonumber 
\frac{1}{T_1T}= \left( \frac{1}{T_1T} \right)_0 + \left( \frac{1}{T_1T} \right)_Q \\
\left( \frac{1}{T_1T} \right)_Q = \frac{\pi k_{\rm B} \gamma^2_n }{(\gamma_e \hbar )^2} \sum_{q \approx Q} A_{hf}^2 \frac{\chi ''_{\perp}(q,\omega)}{\omega},
\end{eqnarray}
where  $\left( 1/T_1T \right)_Q$ is the contribution from wave vectors around $Q$, while $\left( 1/T_1T \right)_0$ denotes the contribution from  $q\sim$ 0, which is proportional to the magnetic susceptibility $\chi_s$. That is, $\left( 1/T_1T \right)_0$ is proportional to the density of states (DOS) at the Fermi level and does not change with temperature. The enhancement of {\tlt} at low temperatures is not caused by impurities, since {\nbr} contains more impurities but the enhancement of {\tlt} is weaker. Therefore, the increase of {\tlt} at low temperatures is ascribed to $\left( 1/T_1T \right)_Q$, i.e. spin correlations develop at low temperatures. On the other hand, since the temperature dependence of the Knight shift is constant above {\tc} in both samples, we conclude that the correlation is of antiferromagnetic nature. The difference in the strength of the antiferromagnetic fluctuations between {\nbr} and {\tar} will be discussed in subsection C in connection with the calculated band structure.

\subsection{Properties of the superconducting state}

\begin{figure}[htbp]
\includegraphics[clip,width=80mm]{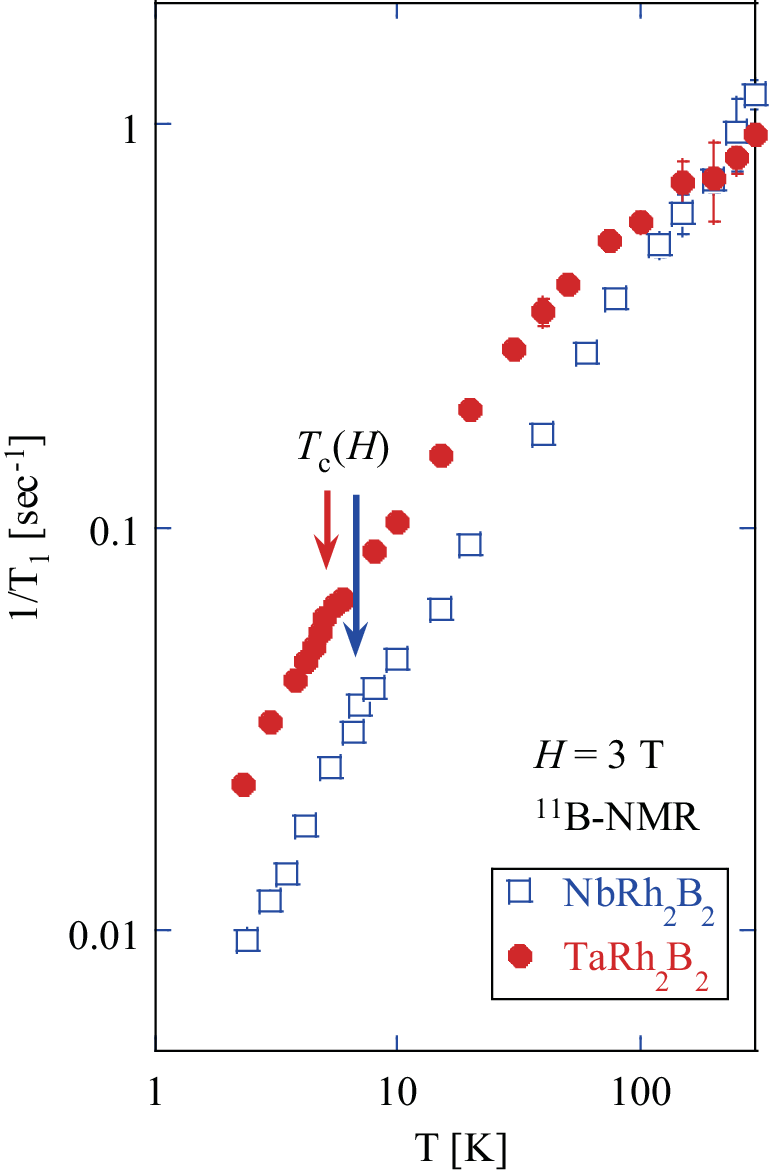}
\caption{\label{T1}(color online) Temperature dependence of the spin-lattice relaxation rate $1/T_1$ measured at the central peak.}
\end{figure}

Figure \ref{T1} shows the temperature dependence of {\tl} measured at the central peak of the NMR spectrum at $H_0$ = 3.0 T. Below {\tc}, {\tl} decreases due to the superconducting transition, but no clear coherence peak was observed for either compound. In BCS superconductors, a large coherence peak is usually observed. A large applied magnetic field could reduce the height of the coherence peak, but the applied magnetic field (3 T) we used is only 1/6 of the upper critical field (18 T) for {\nbr} and 1/4 for {\tar} (11.7 T), which is not strong enough to  completely suppress the coherence peak\cite{tarh2b2_discover,suppress_coherencepeak_doi:10.1143/JPSJ.26.309}.  A large nuclear electric quadrupole moment \cite{Li_PhysRevB.94.174511}, or phonon damping in the strong coupling regime\cite{Maeda_PbTaSe2_PhysRevB.97.184510,Luo_2018} could result in an absence of the coherence peak. However, the former does not apply to $^{11}$B nuclei, and there is no indication of strong-coupling superconductivity as the decrease of {\tl} below {\tc} is not steep. Another candidate to suppress the coherence peak is nonmagnetic impurities. However, in order to suppress completely the coherence peak, one would need 0.7 $n_{\rm cr}$ of impurity, where $n_{\rm cr}$ is the critical impurity concentration to kill superconductivity\cite{griffin}. Therefore, the absence of the coherence peak may indicate the possibility of unconventional superconductivity. In the case of Li$_2$Pd$_3$B and Li$_2$Pt$_3$B, a well-defined coherence peak was found in the former\cite{Nishiyama_PhysRevB.71.220505} but no coherence peak in the latter, which was explained by the different strength of SOC. However, in the present case, no coherence peak is seen in either compound although the SOC is quite different(see section C), which should probably be ascribed to electron correlations. The decrease of {\tl} below {\tc} is not as fast as $T^3$ as seen in, for example, cuprates, which is probably due to impurity scattering\cite{ASAYAMA1991281}. In unconventional superconductors with line nodes in the gap function, impurity scattering can bring about a finite density of states. In fact, when the residual density of states is quite large, a temperature dependence of {\tl} similar to our result was observed\cite{Kawasaki115_2020}.
\begin{figure}[htbp]
\includegraphics[clip,width=80mm]{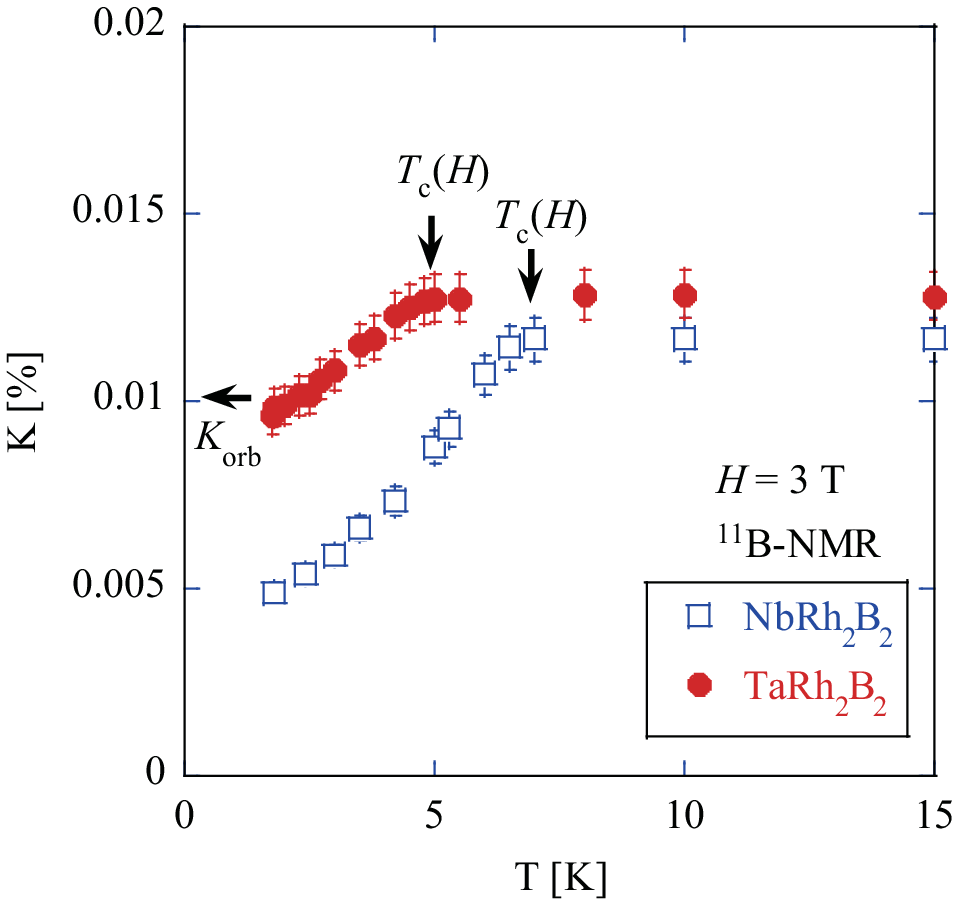}
\caption{\label{Ksc}(color online) Temperature dependence of the Knight shift around {\tc} for {\xr}.
The horizontal arrow indicates the position of $K_{\rm orb}$ (See text).}
\end{figure}

The temperature dependence of the Knight shift around {\tc} is shown in Fig. \ref{Ksc}.
Generally, the Knight shift in the superconducting state is expressed as,
\begin{eqnarray}
K &&= K_{\rm orb}+K_{\rm s} + K_{\rm dia},\\
K_{\rm orb} &&= A_{\rm orb}\chi_{\rm orb} = 2\chi_{\rm orb}\left\langle \frac{1}{r^3} \right\rangle,\\
K_s &&= A_{\rm hf}\chi_{\rm s},\\
\chi_{\rm s} &&= -4\mu^2_{\rm B}\int N_{\rm S}(E)\frac{\partial f(E)}{\partial E}dE,
\end{eqnarray}
where $K_{\rm orb}$ is the contribution due to orbital susceptibility which is $T$-independent, $A_{\rm orb}$ and $A_{\rm hf}$ is the hyperfine coupling constant, $\chi_{\rm orb}$ and $\chi_{\rm s}$ are the orbital and spin susceptibility, and $K_{\rm dia}$ is the contribution from diamagnetism in the vortex state. The $K_{\rm dia}$ is calculated using the following equation for the diamagnetic field $H_{\rm dia}$ \cite{gennes},
\begin{eqnarray}
H_{\rm dia} = H_{\rm c1} \frac{\ln \left(\frac{\beta d}{\sqrt{e}\xi}\right)}{\ln \frac{\lambda}{\xi}}\,.
\end{eqnarray}
Here, $\beta$ is 0.38 for the triangular lattice of the vortex, $\xi$ is the coherence length, $\lambda$ is the London penetration depth. $\xi$ is obtained from the measurement of $H_{\rm c2}$ and $\lambda$ is taken from Ref. \onlinecite{tarh2b2_discover}. 
As a result, $K_{\rm dia}$ was calculated to be -0.025 \% for {\nbr} and -0.015 \% for {\tar},
which are larger than the observed reduction of the Knight shift below {\tc}.

\begin{figure}[htbp]
\includegraphics[clip,width=80mm]{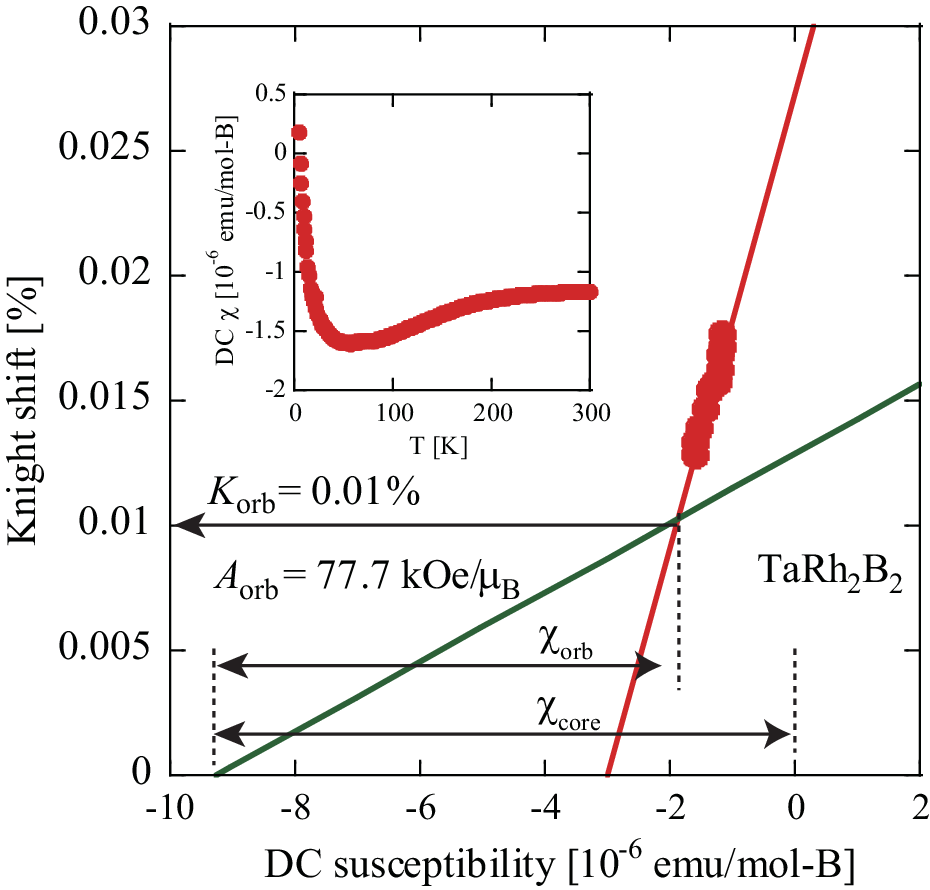}
\caption{\label{K-chi-s}(color online) $K$-$\chi$ plot for {\tar}. The inset shows the results of the dc magnetic susceptibility measurement. The Curie-Weiss-like increase in the DC susceptibility at low temperatures is probably due to paramagnetic impurities.}
\end{figure}

Next, we estimate the $A_{\rm hf}$ and $K_{\rm orb}$ using the $K$-$\chi$ plot. We have performed Dc susceptibility measurements on {\xr}. However, we could not obtain the $K$-$\chi$ plot for {\nbr} due to a dominant Curie-Weiss behavior. In Fig. \ref{K-chi-s}, the Knight shift for the {\tar} is plotted as a function of DC susceptibility. The diamagnetic susceptibility due to closed shells of Ta, Rh, and B was estimated to be $\chi_{core}$ = -1.6$\times$10$^{-6}$ emu$/$mol \cite{Metalic}, from which the slope of $K_{\rm orb} = A_{\rm orb}\chi_{\rm orb}$ was drawn. Here, $\langle 1/r^3 \rangle$=0.62 a.u. is adopted, which is 80\% of the theoretical value for B metal\cite{Atomic}. The orbital part of the shift and susceptibility are $K_{\rm orb} = 0.01\%$, $\chi_{\rm orb} = -9.26 \times 10^{-6}\,{\rm emu}\,/\,{\rm mol}\cdot{\rm B}$. The Knight shift for {\nbr} is reduced below $K_{\rm orb}$, thus most of the decrease can be attributed to the contribution of $K_{\rm dia}$. Therefore, we cannot discuss $K_{\rm s}$ below {\tc} at the moment. The $A_{\rm hf}$ is calculated from the slope of $K$ vs. $\chi$ emu/mol to be $A_{\rm hf}$ =28.7 kOe/$\mu_{\rm B}$.

\subsection{Band structure and other observed properties}

\begin{figure}[htbp]
\includegraphics[width=80mm]{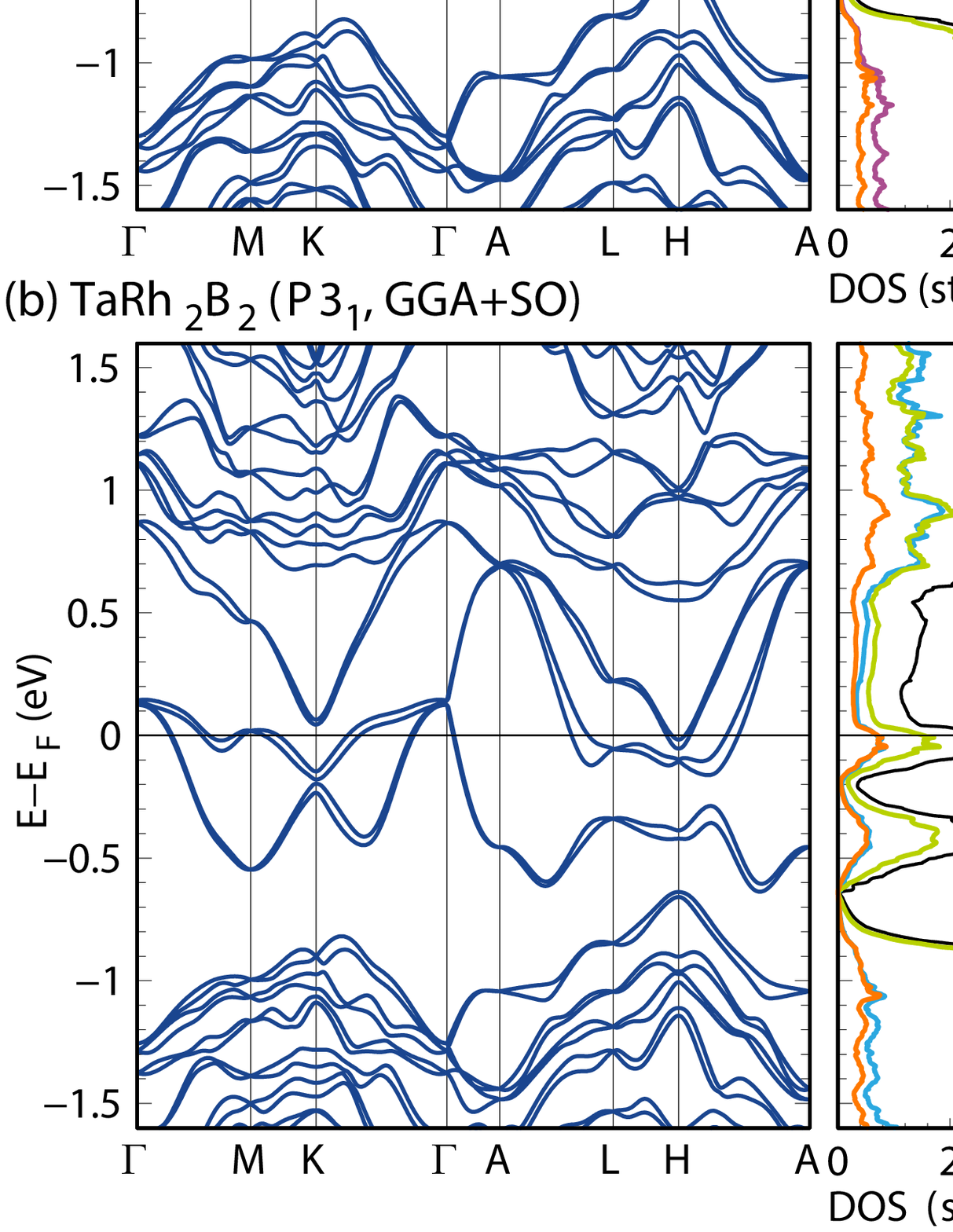}
\caption{\label{DOS}(color online) Band structures and density of states of (a) {\nbr} and (b) {\tar}.}
\end{figure}

In order to understand our results, we performed electronic structure calculations for {\xr}. Figure \ref{DOS} shows fully relativistic GGA bands, labeled as GGA+SO. There is a significant splitting of bands due to spin-orbit coupling; this effect is more pronounced in {\tar} containing heavier Ta compared to {\nbr}. 

We first discuss the differences in {\tc} between {\nbr} (7.8 K) and {\tar} (5.9 K). We evaluated the density of states at the Fermi level $N(E_{\rm F})$. We found $N(E_{\rm F})$ = 3.04564 states/eV/f.u. (f.u. = formula unit) for {\nbr} and $N(E_{\rm F})$ = 2.77660 states/eV/f.u. for {\tar}. Such differences in DOS may be caused by differences in the magnitude of the spin-orbit interaction, namely the magnitude of the band splitting. The $N(E_{\rm F})$ is about 9\% smaller in {\tar}, explaining at least partly why the experimentally observed superconducting {\tc} is also smaller. 

This calculation at first sight seems to contradict with the results of the Knight shift. The Knight shift of {\tar} is larger than that of {\nbr}.
According to Equation (4) to (7), assuming that $K_{\rm orb}$ and $A_{\rm hf}$ are the same for {\xr}, the Knight shift is proportional to the density of states. The difference of the Knight shift indicates that the density of states in {\tar} is larger than in {\nbr}. However, what we observe in $^{11}$B-NMR is the contribution of the $s$-electrons. If we take into account only the $s$-electrons, $N(E_{\rm F})$ is 0.070 states/eV/f.u. for {\nbr} and 0.076 states/eV/f.u. for {\tar} and the contradiction is resolved.

Next, we will discuss effects of the ASOC. For Li$_2$(Pt,Pd)$_3$B, the effect of the ASOC is discussed in terms of band splitting near the Fermi level. Average splitting of the two bands closest to the Fermi level due to SOC is 30 meV in {\nbr} and 50 meV in {\tar}. Spin-orbit interactions and electron correlations are usually in a reciprocal relationship. Namely, in systems with strong electron correlations, SOC is weak and in systems with large SOC, electron correlations are weak. However, there are some exceptions to this principle. Contrary to the usual case, the compound Sr$_2$IrO$_4$ containing a heavy element is a Mott insulator, while Sr$_2$CoO$_4$ and Sr$_2$RhO$_4$ are metals \cite{Sr2CoO4_PhysRevLett.93.167202,Sr2RhO4_PhysRevLett.97.106401,SrIrO4_Kim_PhysRevLett.101.076402,Sr2IrO4_Watanabe_PhysRevLett.105.216410}. Spin-orbit interactions can explain this difference. Sr$_2$IrO$_4$ contains a slightly tilted IrO$_6$ octahedron\cite{Sr2IrO4_Watanabe_PhysRevLett.105.216410}. The stronger ionic nature of the oxide leads to an approximately 5/6 filled $t_{2g}$ shell of Ir. As a result, a Mott insulator is realized. The Mott insulator is realized when spin-orbit interactions and Coulomb interaction further split the orbitals that were initially split by the crystal field in Sr$_2$IrO$_4$\cite{SrIrO4_Kim_PhysRevLett.101.076402}. 

\begin{figure}[htbp]
\includegraphics[width=160mm]{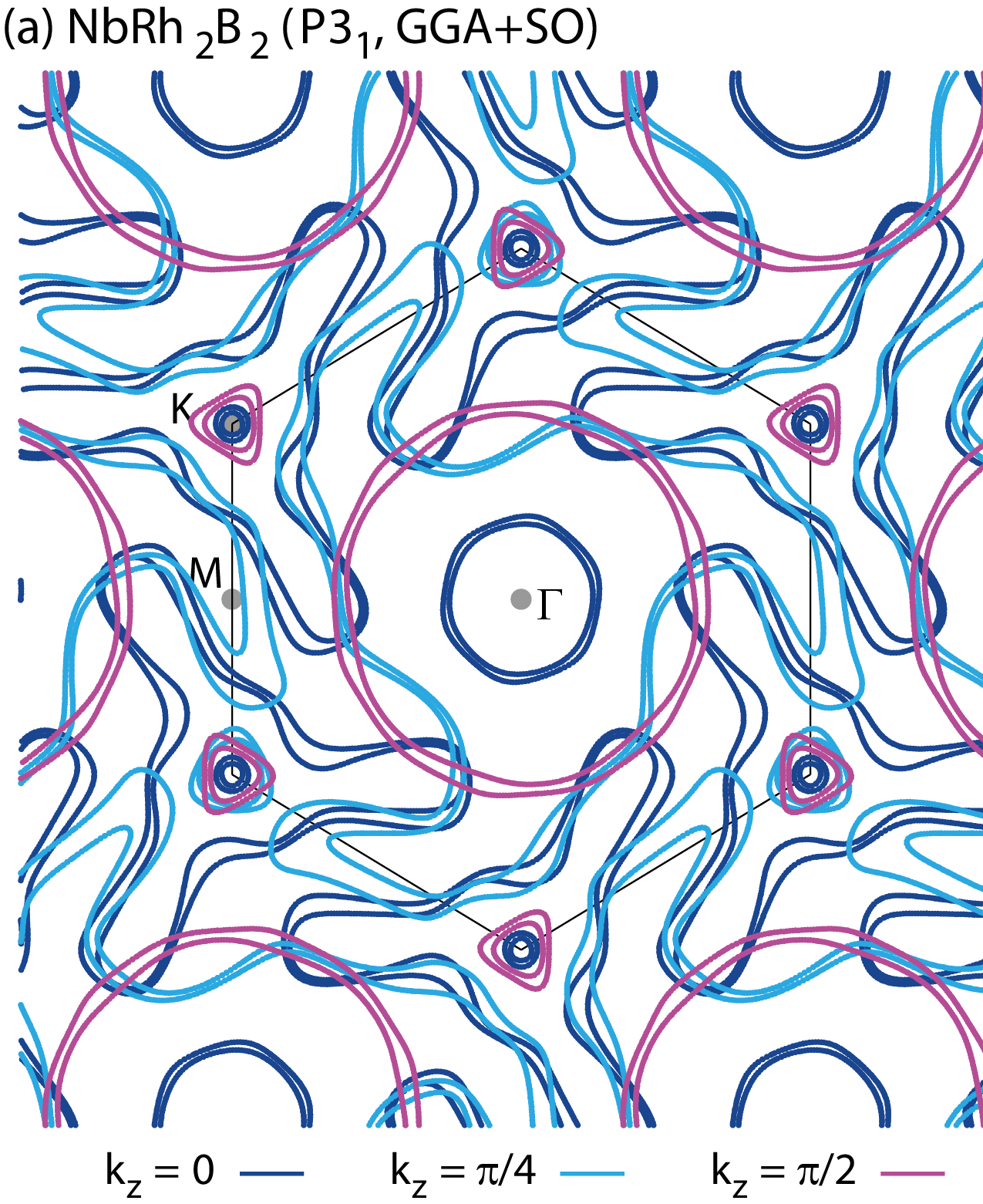}
\caption{GGA+SO Fermi surface cuts of (a) {\nbr} and (b) {\tar}. The
Brillouin zone and high symmetry points at $k_z=0$ are shown.}
\label{fig:fs}
\end{figure}

Similar SOC-assisted electron correlations may be realized in (Nb,Ta)Rh$_2$B$_2$, although {\xr} are intermetallic compounds. Boron tends to form covalent bonds, and since the electronegativities of B, Rh, and Ta/Nb are similar, no substantial transfer of charge is expected. Our calculations of integrated DOS (not shown) support this. Another possibility is the nesting of the Fermi surface. It can be seen that several bands are crossing the Fermi level. The spin-orbit interaction may have caused a different Fermi level nesting between the two materials. As mentioned above, the SOC is larger in {\tar}. We calculated the electronic structure of {\nbr} and {\tar} using $200\times 200$ $k$ meshes and the GGA+SO functional in order to extract the Fermi surfaces at three different $k_z$ values. The result is shown in Fig.\ref{fig:fs}. The splitting of bands due to SOC is clearly stronger in {\tar}, and the Fermi surface nesting is quite different for the two compounds. For example, the six-fold symmetric Fermi surface for {\tar} appears to have a better nesting condition compared to {\nbr}.

\section{summary}
In summary, we have performed $^{11}$B-NMR measurements on the non-centrosymmetric superconductors {\nbr} and {\tar} with chiral structure. In both compounds, we found that {\tlt} increases with decreasing $T$ at low temperatures. On the other hand, the Knight shift was constant at low temperatures for both compounds. These results point to the existence of antiferromagnetic spin correlations. Furthermore, the magnitude of the spin correlation is much more significant for {\tar} than {\nbr}. It is not usual for compounds containing heavy elements to have stronger spin correlations. It is necessary to consider the possibility that spin-orbit interactions enhance spin correlations in this system. We hope that our result will stimulate more works in this direction. In the superconducting state, no coherence peak was observed in {\tl} just below {\tc}, suggesting unconventional superconductivity.
\begin{acknowledgments}
This work was supported by Research Grants from MEXT and JSPS, No. JP19H00657, No. JP19K03747,  and No. JP20K03862.
We acknowledges stimulating discussions with I. I. Mazin.
\end{acknowledgments}

\bibliography{TRh2B2}

\end{document}